\documentclass[a4paper]{jpconf}
\usepackage{graphicx}
\usepackage{wrapfig}
\usepackage{amsmath,amssymb}

\newcommand{\bra}[1]{\langle {#1} |}
\newcommand{\ket}[1]{| {#1} \rangle}
\newcommand{\inproduct}[2]{\langle #1 | #2 \rangle}

\begin{document}
\title{
Density functional approaches to nuclear dynamics
}

\author{
T.~Nakatsukasa$^{1,2}$,
S.~Ebata$^{1,3}$,
P.~Avogadro$^{1,4}$,
L. Guo$^{1,5}$,
T.~Inakura$^1$ and
K.~Yoshida$^6$
}

\address{
$^1$ RIKEN Nishina Center, 2-1 Hirosawa, Wako-shi, 351-0198, Japan\\
$^2$ Center for Computational Sciences, University of Tsukuba, Tsukuba 305-8571, Japan\\
$^3$ Center for Nuclear Study, University of Tokyo, Wako-shi 351-0198, Japan\\
$^4$ Department of Physics \& Astronomy, Texas A\&M
University-Commerce, Commerce, Texas 75428, USA\\
$^5$ College of Physical Sciences, Graduate University of Chinese Academy
of Sciences, Beijing 100049, China\\
$^6$ Department of Physics, Niigata University, Niigata 950-2181, Japan
}

\begin{abstract}
We present background concepts of the nuclear density functional theory (DFT)
and applications of the time-dependent DFT with the Skyrme energy
functional for nuclear response functions.
Practical methods for numerical applications of
the time-dependent Hartree-Fock-Bogoliubov theory (TDHFB) are proposed;
finite amplitude method and canonical-basis TDHFB.
These approaches are briefly reviewed and some numerical applications
are shown to demonstrate their feasibility.
\end{abstract}

\section{Introduction}

The nucleus is a quantum object.
The nuclear interaction is not strong enough to 
localize the nucleonic wave function, thus, the nuclear matter stays
in the liquid phase even at zero temperature \cite{BM69}.
This strong quantum nature in nuclei leads to a rich variety
of unique phenomena.
Extensive studies have been made for
constructing theoretical models to elucidate basic nuclear dynamics behind
a variety of nuclear phenomena.
Simultaneously, significant efforts have been made in the microscopic
foundation of those models.

Although there have been significant developments in 
the ``first-principles'' large-scale computation,
starting from the bare nucleon-nucleon (two-body \& three-body) forces,
they are still limited to light nuclei with small mass numbers,
typically $A\lesssim 10$ \cite{PW01}.
In contrast, the density functional theory (DFT) is currently
a leading theory for describing
nuclear properties of heavy nuclei \cite{BHR03,LPT03}.
It is capable of
describing almost all nuclei, including nuclear matter,
with a single universal energy density functional (EDF).
In addition, its strict theoretical foundation is given by
the basic theorem of the DFT \cite{HK64,KS65}.
Since the nucleus is a self-bound system without an external
potential, the DFT theorem should be modified from its original form.
This problem was addressed by recent studies \cite{Eng07,Gir08,GJB08}.
In this paper, we present basic concepts of the EDF in nuclei
and feasible methodologies of the time-dependent DFT.

\section{Backgrounds of nuclear energy density functionals}
\label{sec: basic_concepts_EDF}

In this section, basic properties of nuclei and
historical developments in nuclear structure theory,
which leads to the nuclear energy
density functional, are briefly reviewed.
To simplify the discussion here, we consider an infinite
uniform nuclear matter neglecting the Coulomb interaction.

\subsection{Basic property of nuclear systems: Saturation and
 independent-particle motion}
\label{sec:basic_property}

\subsubsection{Saturation}
The volume and total binding energy of
observed nuclei in nature are approximately
proportional to the mass number $A$.
In other words, they have
an approximately constant density $\rho_0\approx 0.17$ fm$^{-3}$
and a constant binding energy per particle $B/A\approx 8$ MeV.
Thus, extrapolating this property to the infinite nuclear matter
neglecting the Coulomb interaction,
the nuclear matter should have an equilibrium state with
$\rho_0\approx 0.17$ fm$^{-3}$ and $B/A\approx 16$ MeV,
at zero pressure and zero temperature.
This property are called ``saturation property'',
that is analogous to the liquid.
The most famous and successful model based on this liquid picture
of nuclei is the empirical mass formula of Bethe and Weizs\"acker
\cite{Wei35,BB36}.
This formula contains the surface and Coulomb terms in addition to
the leading term proportional to $A$, which well accounts of
the bulk part of the nuclear binding.

\subsubsection{Independent-particle motion in nuclei}

There are many evidences for the fact that
the mean-free path of nucleons
is larger than the size of nucleus.
The great success of the nuclear shell model \cite{MJ55} gives
such an example,
in which nucleons are assumed to move independently inside an
average one-body potential.
The scattering experiments with incident neutrons and protons
provide more quantified information on the mean-free path.
In fact, the mean free path depends on the nucleon's energy,
and becomes larger for lower energy \cite{BM69}.
Therefore, it is natural to assume that
the nucleus can be primarily approximated by the
independent-particle model with an average one-body potential.
For the nuclear matter,
this approximation leads to the degenerate Fermi gas of the same number of
protons and neutrons ($Z=N=A/2$).
The observed saturation density of $\rho_0\approx 0.17$ fm$^{-3}$
gives the Fermi momentum, $k_F\approx 1.36$ fm$^{-1}$,
which corresponds to the Fermi energy (the maximum kinetic energy),
$T_F=k_F^2/2M\approx 40$ MeV.

\subsection{Problems of a mean-field picture}
\label{sec:IPM}

Evidences of the independent-particle motion
encourage us to adopt the mean-field picture of nuclei.
However, it turned out that the mean-field models cannot describe
the nuclear saturation property.
Let us explain this for the uniform nuclear matter
with a constant attractive potential $V<0$.

The constancy of $B/A$ means that it is equal to the
separation energy of nucleons, $S$.
In the Fermi-gas model, it is estimated as
\begin{equation}
\label{B1}
S \approx B/A \approx -(T_F + V) .
\end{equation}
Since the binding energy is $B/A\approx 16$ MeV,
the potential $V$ is about $-55$ MeV.
It should be noted that the relatively small separation energy is
the consequence of the significant cancellation between
kinetic and potential energies.
In the mean-field theory,
the total (binding) energy is given by
\begin{equation}
\label{B2}
-B = \sum_{i=1}^A \left( T_i + \frac{V}{2} \right)
        = A \left( \frac{3}{5}T_F + \frac{V}{2} \right) ,
\end{equation}
where we assume that the average potential results from a two-body
interaction.
The two kinds of expressions for $B/A$, Eqs. (\ref{B1}) and (\ref{B2}),
lead to $T_F\approx -5V/4\approx 70$ MeV,
which is different from the previously estimated value ($\sim 40$ MeV).
Moreover,  the negative separation energy ($T_F +V >0$)
contradicts the fact that the nucleus is bound!

To reconcile
the independent-particle motion with the saturation property of the nucleus,
the nuclear average potential must be state dependent.
Allowing the potential $V_i$ depend on the state $i$,
the potential $V$ should be replaced by that for the highest occupied
orbital $V_F$ in Eq. (\ref{B1}),
and by its average value $\langle V \rangle$ in
the right-hand side of Eq. (\ref{B2}).
Then, we obtain the following relation:
\begin{equation}
\label{V_F}
V_F \approx \langle V \rangle + T_F/5 + B/A .
\end{equation}
Therefore, the potential $V_F$ is shallower
than its average value.

Weisskopf suggested the momentum-dependent potential $V$, which can be
expressed in terms of an effective mass $m^*$ \cite{Wei57}:
\begin{equation}
\label{mom_dep_pot}
V_i=U_0+U_1\frac{k_i^2}{k_F^2} .
\end{equation}
Actually, if the mean-field potential is non-local, it can be
expressed by the momentum dependence.
Equation (\ref{mom_dep_pot})
leads to the effective mass, $m^*/m = (1+U_1/T_F)^{-1}$.
Using Eqs. (\ref{B1}), (\ref{V_F}), and (\ref{mom_dep_pot}),
we obtain the effective mass given by
\begin{equation}
\label{m*/m}
\frac{m^*}{m} = \left\{ \frac{3}{2} + \frac{5}{2}\frac{B}{A}\frac{1}{T_F}
 \right\}^{-1} \approx 0.4 .
\end{equation}
Quantitatively, this value disagrees with the experimental data.
The empirical values of the effective mass
vary according to the energy of nucleons,
$0.7 \lesssim m^*/m \lesssim 1$,
however, they are almost twice larger than
the value in Eq. (\ref{m*/m}).
As far as we use a normal two-body interaction,
this discrepancy should be present in the mean-field calculation
with any interaction,
because Eq. (\ref{m*/m}) is valid in general
for a saturated self-bound system.
Therefore, the naive mean-field models
have a fundamental difficulty to describe the nuclear saturation.

\subsection{Nucleon-nucleon interaction (nuclear force)}
\label{sec:NN}

To understand the origin of the problem, properties of the nuclear
force provides an important key.
The saturation property of nuclear density reflects a balance
between attractive and repulsive contributions to nuclear binding
energy.
One source of such repulsive effects is the nucleonic kinetic energy
of the Fermi gas.
However, its contribution per particle is proportional to $\rho^{2/3}$,
which is not strong enough to resist against the collapse caused by
the attractive force between nucleons.
Therefore, the nucleonic interaction must contain a repulsive element.
Indeed, the phase-shift analysis on the nucleon-nucleon scattering
at high energy ($E_{\rm lab}>250$ MeV) reveals a short-range strong
repulsive core in the nucleonic force.
The radius of the repulsive core is approximately $c\approx 0.5$ fm.
This strong repulsive core prevents the nucleons approaching
closer than the distance $c$, which produces a strong two-body correlation,
$\rho^{(2)}(\vec{r}_1,\vec{r}_2)\approx 0$ for $|\vec{r}_1-\vec{r}_2|<c$.
The attractive part of the interaction has a longer range, which can
be characterized by the pion's Compton wave length $\lambda_\pi$,
and is significantly weaker than the repulsion.
Thus, a naive application of the mean-field calculation fails to bind
the nucleus, since the mean-field approximation cannot take account
of such strong two-body correlations.

At first sight,
this seems inconsistent with the experimental observations.
As we mentioned in Sec~\ref{sec:basic_property},
there are many experimental evidences for the independent-particle
motion in nuclei.
We may intuitively understand that it is due to the fact that the
nucleonic density is significantly smaller than $1/c^3$.
Therefore, the collisions by the repulsive core rarely occur and
the system can be approximately described in terms of the
independent-particle motion.
Furthermore, the effects of the Pauli principle hinder the collisions,
since the nucleons cannot be scattered into occupied states.
Although the repulsive-core collisions are experienced by only
a small fraction of nucleons ($\sim \rho_0 c^3$),
each collision carries a large amount of energy.
Therefore, the repulsive core provides an important contribution to the
total energy and are responsible for the saturation.

Another important factor for the independent particle motion
is the strong quantum nature due to the weakness of the attractive
part of the nuclear force.
The importance of the quantum nature can be measured by the
magnitude of the zero-point kinetic energy compared to that of the
interaction.
If the attractive part of the nuclear force were much stronger
than the unit of $\hbar^2/Mc^2$,
the quantum effect would disappear and
each nucleon would stay at the bottom of the interaction potential.
Then, the nucleus would crystallize at low temperature.
In reality, the attraction of the nuclear force is so weak
that it barely produces many-nucleon bound states at the relatively
low density.

\subsection{From Brueckner theory to EDF}
\label{sec:DDHF}

The nuclear matter theory pioneered by Brueckner
gives a hint for a solution for the previous difficulty to
understand the nuclear saturation.
The Brueckner theory may provide a first step toward the quantitative
treatment to understand the saturation property and the
independent-particle motion in nuclei.
Details of the theory can be found at Refs. \cite{Day67,RS80}.

The basic ingredient of the Brueckner theory is a two-body scattering
matrix of particle 1 and 2 inside nucleus caused by the nuclear force $v$,
\begin{equation}
\label{G-matrix}
G(\omega)\equiv v+v \frac{Q}{\omega-Q(T_1+T_2)Q} G(\omega),
\end{equation}
where $T_i$ is the kinetic energy of particle $i$,
$Q$ is the Pauli-exclusion operator to restrict the intermediate states,
and $\omega$ is called a starting energy that depends on energies of
particle 1 and 2.
This is called $G$-matrix \cite{BG57}.
The $G$-matrix renormalizes high-momentum components in the bare nuclear
force and becomes an effective interaction in nuclei under
the independent-pair approximation.
The $G$-matrix reflects an underlying structure
of the independent many-nucleon system through the operator $Q$ and the
starting energy $\omega$.
Inevitably, the $G$-matrix becomes state (structure) dependent.

Since the short-range singularity is renormalized in the $G$-matrix,
we can calculate the total energy in the independent-particle
(mean-field) model, analogous to Eq. (\ref{B2}).
\begin{equation}
-B=\sum_{i=1}^A \left\{ T_i +
 \frac{1}{2} \sum_{j=1}^A
\bar{G}_{ij,ij}(\omega_{ij})
\right\}
\end{equation}
where $\omega_{ij}=\epsilon_i+\epsilon_j$,
defines the self-consistency condition for the Brueckner's
single-particle energies,
and $\bar{G}_{ij,ij}\equiv G_{ij,ij}-G_{ij,ji}$.
This is called Brueckner-Hartree-Fock (BHF) theory.
The validity of the BHF theory is measured by the wound integral
$\kappa=\inproduct{\psi-\phi}{\psi-\phi}$, where $\ket{\phi}$ is
an unperturbed two-particle wave function and $\ket{\psi}$ is
a correlated two-particle wave function in nucleus.
$\kappa$ is known to be about 15 \%.
The BHF calculation was successful to describe the nuclear saturation,
however, could not simultaneously reproduce empirical values of
$B/A$ and $\rho_0$,
known as a problem of the Coester band \cite{Day81a}.
Its applications to finite nuclei also have similar problems to reproduce
the energy, radius, and density in the ground state.

A possible solution to these problems was given by Negele \cite{Neg70}.
Starting from a realistic $G$-matrix, using the expressions for
the Pauli operator
\begin{equation}
\bra{\vec{r}_1\vec{r}_2}Q\ket{\vec{r}'_1\vec{r}'_2}
= \left\{ \delta(\vec{r}_1-\vec{r}'_1) - \rho(\vec{r}_1-\vec{r}'_1) \right\}
\left\{ \delta(\vec{r}_2-\vec{r}'_2) - \rho(\vec{r}_2-\vec{r}'_2) \right\} ,
\end{equation}
and the average single-particle energy $\epsilon[\rho(\vec{r})]$,
the local density approximation is introduced to expand the off-diagonal
density matrix $\rho(\vec{r}+\vec{s}/2,\vec{r}-\vec{s}/2)$
with respect to the relative coordinate $|\vec{s}|$.
Then, a short-range part of the $G$-matrix, which is not fully understood,
is phenomenologically added to the
energy expression to quantitatively fit the saturation property,
and finally, the total energy is treated variationally.
This procedure is called the density matrix expansion (DME) \cite{NV72}.
The state dependence of the $G$-matrix is now replaced by the
density dependence.
The final result for the total energy, for a uniform nuclear matter,
is written as a function of the neutron and proton densities,
$\rho_n$ and $\rho_p$, and the kinetic densities,
$\tau_n$ and $\tau_p$;
$E=E[\rho_n,\rho_p,\tau_n,\tau_p]$.
This expression can be generalized for finite nuclei as
\begin{equation}
E= \int d\vec{r} H(\vec{r}) ,
\quad H(\vec{r})=H[\rho_n(\vec{r}),\rho_p(\vec{r}),\tau_n(\vec{r}),\tau_p(\vec{r})] ,
\end{equation}
$H(\vec{r})$ here is
a complete analogue of the Hamiltonian density in the
Skyrme EDF (without the spin-orbit and Coulomb
terms) \cite{VB72}.
Thus, the DME with a microscopic G-matrix leads to a variational
treatment of a simple EDF of the Skyrme type.

The essential aspect of DME is in its non-trivial density dependence
and the variational treatment.
Now, the expression for the total energy, Eq. (\ref{B2}), should be modified
due to these non-trivial density-dependent terms.
This resolves the previous issue, and provides a consistent
independent-particle description for the nuclear saturation.
In nuclear physics, this was often interpreted
as the density-dependent effective interaction.
In this terminology,
the variation of the total energy with respect to the density
contains re-arrangement potential, $\partial V_{\rm eff}[\rho]/\partial\rho$,
which comes from the density dependence of the effective force
$V_{\rm eff}[\rho]$.
These terms turn out to be crucial to obtain the saturation condition.

In summary, the failure in the mean-field description of nuclei
using phenomenological effective interactions can be traced back
to the missing state (structure) dependence.
The EDF approaches take into account the state dependence
in terms of the non-trivial density dependence.

\section{Time-dependent Hartree-Fock-Bogoliubov theory}
\label{sec: TDKSB}

For a quantitative description of heavy nuclei in open-shell configurations,
it is necessary to include the pairing correlations.
This can be done by a straightforward extension of the
Skyrme energy functional simply added by the pairing energy functional.
\begin{equation}
E[\rho,\kappa]=E_{\rm Skyrme}[\rho] + E_{\rm pair}[\rho,\kappa] ,
\end{equation}
where $\kappa$ is the pair density.
The variation of the energy with the constraint on the average
particle number leads to the Hartree-Fock-Bogoliubov (HFB)
equation \cite{RS80,BR86}.
However, here, let us start from the time-dependent Hartree-Fock-Bogoliubov
(TDHFB) equation,
\begin{equation}
\label{TDHFB_equation}
i\frac{\partial}{\partial t} {\cal R}(t) = \left[ {\cal H}(t) ,{\cal R}(t) \right] ,
\end{equation}
where ${\cal H}(t)$ is the HFB Hamiltonian usually written in a form \cite{RS80}
\begin{equation}
\label{HFB_Hamiltonian}
{\cal H}(t)= {\cal H}[{\cal R}(t)] \equiv 
\begin{pmatrix}
h[\rho(t),\kappa(t)] & \Delta[\rho(t),\kappa(t)] \\
-\Delta^*[\rho(t),\kappa(t)] & -h^*[\rho(t),\kappa(t)]
\end{pmatrix}
.
\end{equation}
The generalized density matrix $R(t)$ is given by
\begin{equation}
{\cal R}(t)=
\begin{pmatrix}
\rho(t) & \kappa(t) \\
-\kappa^*(t) & 1-\rho^*(t)
\end{pmatrix}
\end{equation}
which is Hermitian and idempotent: ${\cal R}^2={\cal R}$.
Thus, the eigenvalues of ${\cal R}(t)$ are 0 and 1.
The eigenvectors
$\Psi_\nu(t)=\begin{pmatrix} U_\nu(t) \\ V_\nu(t)\end{pmatrix}$,
which correspond to the eigenvalue 0, are called
``unoccupied'' quasiparticle (qp) orbitals.
For every unoccupied qp orbital, there exits a conjugate ``occupied'' orbital
$\bar\Psi_\nu(t)=\begin{pmatrix} V_\nu^*(t) \\ U_\nu^*(t)\end{pmatrix}$,
whose eigenvalue is 1.

We define the following matrix of size $2M\times M$ where $M$ is
the dimension of the adopted single-particle Hilbert space,
\begin{equation}
\label{Psi_matrix}
\Psi_{\alpha\nu}(t)=\begin{cases}
\inproduct{\alpha}{U_\nu(t)} & \alpha=1,\cdots,M \\
\inproduct{\alpha-M}{V_\nu(t)} & \alpha=M+1,\cdots,2M
\end{cases}
\end{equation}
which collectively represents time-dependent unoccupied qp orbitals
($\nu=1,\cdots,M$).
The matrix for the occupied orbitals $\bar\Psi(t)$
are defined in the same manner,
with $\Psi_\nu(t)$ replaced by $\bar\Psi_\nu(t)$.
Using these matrices, ${\cal R}(t)$ can be written in a simple form as
${\cal R}(t)=\bar\Psi(t)\bar\Psi^\dagger(t) =
1-\Psi(t)\Psi^\dagger(t)$.
The orthonormal property of the qp orbitals is given by
$\Psi^\dagger(t) \Psi(t) = \bar\Psi^\dagger(t)\bar\Psi(t)=1$.
Note that ${\cal R}(t)$ can be regarded as the projection operator
onto the occupied space.
Substituting these expressions into Eq. (\ref{TDHFB_equation}),
we obtain the TDHFB equation for the qp orbitals as
\begin{equation}
\label{G-TDKSB}
{\cal R}(t) \left\{i\frac{\partial}{\partial t} - {\cal H}(t)\right\}\Psi(t)
= 0 ,
\quad\quad
{\cal Q}(t) \left\{i\frac{\partial}{\partial t} - {\cal H}(t)\right\}\bar\Psi(t)
= 0 ,
\end{equation}
where ${\cal Q}(t)=1-{\cal R}(t)$ is the projection operator onto
the unoccupied space.
Therefore, the occupied-occupied and unoccupied-unoccupied
matrix elements of
$i\partial/\partial t -{\cal H}(t)$ are arbitrary at every instant of time.
This is related to the $U(M)$ invariance;
${\cal R}(t)$ is invariant with respect to the transformation
among the occupied (unoccupied) qp orbitals,
$\bar\Psi' = \bar\Psi U$ ($\Psi' = \Psi U$), where $U$ is
an $M\times M$ unitary matrix.

Now, let us derive the static Hartree-Fock-Bogoliubov (HFB) equation
as a {\it quasi-stationary} solution of Eq. (\ref{TDHFB_equation}).
This does not correspond to the solution for
$\partial {\cal R}/\partial t = 0$,
because the solution must have a given value of the average particle number;
${\rm tr}\rho=N$.
When the system is in the superfluid phase ($\kappa\neq 0$),
${\cal R}(t)$ and ${\cal H}(t)$ do not commute with the matrix associated
with the particle number, ${\cal N}\equiv
\begin{pmatrix}
1 & 0 \\ 0 & -1
\end{pmatrix}
$.
Thus, we may construct the generalized density $R(t)$ defined
in the frame of reference
rotating in the gauge space.
\begin{equation}
R(t) =\exp(i\mu{\cal N}t) {\cal R}(t) \exp(-i\mu{\cal N}t)
=
\begin{pmatrix}
\rho(t) & \kappa(t) e^{2i\mu t} \\
-\kappa^*(t) e^{-2i\mu t} & 1-\rho^*(t)
\end{pmatrix}
.
\end{equation}
For this $R(t)$, TDHFB equation (\ref{TDHFB_equation}) becomes
\begin{equation}
\label{TDKSB_equation_2}
i\frac{\partial}{\partial t} R(t) =
\left[ H(t)-\mu {\cal N}, R(t) \right] ,
\end{equation}
where
\begin{equation}
H(t) = 
\exp(i\mu{\cal N}t) {\cal H}(t) \exp(-i\mu{\cal N}t)
=\begin{pmatrix}
h(t) & \Delta(t) e^{2i\mu t} \\
-\Delta^*(t) e^{-2i\mu t} & -h^*(t)
\end{pmatrix}
\end{equation}
Namely, the transformation does not change $\rho$ and $h(t)$, but modifies
$\kappa(t)$ and $\Delta(t)$ by multiplying the time-dependent phase
$e^{2i\mu t}$.
As far as $\Delta(t)$ linearly depends on $\kappa(t)$ and
$h(t)$ is a functional of a product of $\kappa(t)$ and $\kappa^*(t)$,
$H[R(t)]$ has the functional form identical to ${\cal H}[{\cal R}(t)]$.
The stationary state can be defined by the time-independent density
in the rotating frame, $\partial R/\partial t=0$, leading to
\begin{equation}
\left[ H-\mu {\cal N}, R \right] = 0 ,
\end{equation}
where $\mu$ correspond to the chemical potential that should be
determined by the condition on the particle number.
With a proper choice of the $U(M)$ degrees of freedom,
the quasiparticle orbitals $\Phi_\nu$ become time-independent eigenstates as
$(H-\mu {\cal N})\Phi_\nu=E_\nu\Phi_\nu$.
The collective representations, $2M\times M$ matrices $\Phi$ and $\bar\Phi$,
are constructed in the same manner as Eq. (\ref{Psi_matrix}).

The HFB state is a {\it quasi-stationary} state in which
the pair density and potential are {\it rotating} in the complex plain
with a constant angular velocity $2\mu$ . 
This is a collective motion corresponding to the Nambu-Goldstone mode
associated with the spontaneous breaking
of the gauge symmetry, called ``pair rotation''.

\section{Finite amplitude method}

Although there have been significant efforts to develop a numerical
solver of the TDHFB equation in last decade \cite{HN07,SBMR11},
the computation of the full three-dimensional (3D) dynamics
is still a difficult and challenging task.
A possible simplification is given by the small-amplitude approximation.
It is well-known that this will lead to the quasiparticle random-phase
approximation (QRPA).
Recently, we have proposed a feasible approach for the linear response
calculation \cite{NIY07} and it has been extended for the QRPA \cite{AN11}.
The method is called ``finite amplitude method'' (FAM).

The QRPA equations can be written in a form
\begin{equation}
\label{LRE_QRPA}
\begin{split}
(E_\nu + E_{\nu'}-\omega ) X_{\nu\nu'}(\omega) + (H_1)^{20}_{\nu\nu'}(\omega)
 = -({\cal V}_1)^{20}_{\nu\nu'}(\omega) , \\
(E_\nu + E_{\nu'}+\omega ) Y_{\nu\nu'}(\omega) + (H_1)^{02}_{\nu\nu'}(\omega)
 = -({\cal V}_1)^{02}_{\nu\nu'}(\omega) ,
\end{split}
\end{equation}
where
\begin{eqnarray}
(H_1)^{20}_{\nu\nu'}(\omega) &=&
  \left[ \Phi^\dagger H_1(\omega) \bar\Phi \right]_{\nu\nu'} , \quad
 ({\cal V}_1)^{20}_{\nu\nu'}(\omega) =
  \left[ \Phi^\dagger {\cal V}_1(\omega) \bar\Phi \right]_{\nu\nu'} , \\
(H_1)^{02}_{\nu\nu'}(\omega) &=&
  -\left[ \bar\Phi^\dagger H_1(\omega) \Phi \right]_{\nu\nu'} , \quad
 ({\cal V}_1)^{02}_{\nu\nu'}(\omega) =
  -\left[ \bar\Phi^\dagger {\cal V}_1(\omega) \Phi \right]_{\nu\nu'} .
\end{eqnarray}
Here, ${\cal V}_1(\omega)$ is an external field and $H_1(\omega)$ 
is an induced residual field.
The most tedious part is the calculation of
$H_1(\omega)$ expanded in terms of $X(\omega)$ and $Y(\omega)$.
The FAM calculates the induced residual fields as follows \cite{AN11}:
\begin{equation}
H_1^{20}(\omega) =  \Phi^\dagger
\frac{H[R_\eta] - H[R_0]}{\eta}
\bar\Phi
, \quad\quad
H_1^{02}(\omega) =  -\bar\Phi^\dagger
\frac{H[R_\eta] - H[R_0]}{\eta}
\Phi .
\end{equation}
where $\eta$ is a small real parameter.
Here,
$R_\eta(\omega) 
               = \bar\Psi'_\eta(\omega) \bar\Psi_\eta^\dagger(\omega)
$,
with
\begin{equation}
\bar\Psi'_\eta(\omega) = \bar\Phi + \eta \Phi X(\omega) ,\quad\quad
\bar\Psi_\eta^\dagger(\omega) = \left( \bar\Phi + \eta \Phi Y^*(\omega) \right)^\dagger.
\end{equation}
Equivalently, the FAM formula can be written in terms of the qp orbitals as
\begin{eqnarray}
H_1^{20}(\omega) =  \Phi^\dagger
\frac{H[\bar\Psi_\eta',\bar\Psi_\eta^\dagger]
 - H[\bar\Phi,\bar\Phi^\dagger]}{\eta}
\bar\Phi ,
\quad\quad
H_1^{02}(\omega) =  -\bar\Phi^\dagger
\frac{H[\bar\Psi_\eta',\bar\Psi_\eta^\dagger]
 - H[\bar\Phi,\bar\Phi^\dagger]}{\eta}
\Phi .
\end{eqnarray}
\begin{wrapfigure}{r}{0.5\textwidth}
\vspace{-10pt}
\includegraphics[width=0.5\textwidth]{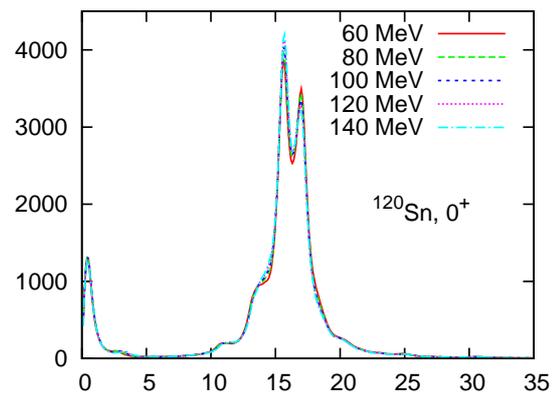}
\caption{Isoscalar monopole strength function in $^{120}$Sn.
See text for details.}
\label{fig: 120Sn}
\end{wrapfigure}
According to these FAM formulae, the residual fields can be calculated
by using the finite difference with the parameter $\eta$.
In this way, the calculation requires only the HFB Hamiltonian
$H[R]$.

Computer programs of the FAM have been developed for spherical
nuclei \cite{AN11} based on the {\sc hfbrad} \cite{hfbrad},
and for axially deformed nuclei \cite{Sto11} based on the {\sc hfbtho}
\cite{hfbtho}.
The FAM in the 3D grid-space representation was also achieved
\cite{INY09,INY11} for nuclei in the normal phase.
Here, we show the isoscalar monopole strength function in $^{124}$Sn
in Fig. \ref{fig: 120Sn}.
The ground state is obtained by the {\sc hfbrad} using the SkM*
functional and the volume-type pairing.
The zero-range nature of the pairing functional requires us to truncate
the model space.
The truncation was done according to the cutoff for the qp energy,
and results with different cutoff energies are shown
in Fig. \ref{fig: 120Sn}  as well.
The final results are almost independent from the choice of the cutoff.

\section{Canonical-basis TDHFB}

In this section, we present another approximate treatment of
the TDHFB.
Here, we do not take the small amplitude limit, instead,
assume the diagonal property of the pair potential.
In fact, in the stationary limit, this corresponds to the
the well-known BCS approximation for the ground state \cite{RS80}.
Therefore, it can be regarded as the BCS-like approximation
in the time-dependent treatment.

At every instant of time, we may identify the canonical basis
in which the density is diagonal,
$\rho_{ij}(t)=\rho_{i}(t)\delta_{ij}$.
The canonical states $(i,\bar{i})$ are always paired with
$\rho_{ii}(t)=\rho_{\bar{i}\bar{i}}(t)=\rho_{i}(t)$ and
$\kappa_{i\bar{j}}(t)=\kappa_i(t) \delta_{ij}$.
Here, we introduce an assumption that the pair potential
is also diagonal in the canonical basis,
$\Delta_{i\bar{j}}=-\Delta_i(t) \delta_{ij}$
In the stationary limit,
this approximation is identical to the usual BCS approximation \cite{RS80}.
Then, we end up the following set of equations \cite{Eba10}.
\begin{eqnarray}
\label{Cb-TDHFB-1}
&& i\frac{\partial}{\partial t} \ket{i(t)} = \left\{ h(t) - \eta_i(t) \right\}
\ket{i(t)} ,\quad\quad
i\frac{\partial}{\partial t} \ket{\bar{i}(t)} = \left\{ h(t) - \eta_{\bar{i}}(t) \right\}
\ket{\bar{i}(t)} , \\
\label{Cb-TDHFB-2}
&& i\frac{d}{dt} \rho_i(t) = \kappa_i(t) \Delta_i^*(t) 
                        - \kappa_i^*(t) \Delta_i(t) , \\
\label{Cb-TDHFB-3}
&& i\frac{d}{dt} \kappa_i(t) = \left\{ \eta_i(t) + \eta_{\bar{i}}(t) \right\}
                      \kappa_i(t) + \Delta_i(t) 
                     \left\{ 2\rho_i(t) -1 \right\} .
\end{eqnarray}
Here, $\eta_i(t)$ and $\eta_{\bar{i}}(t)$ are arbitrary real functions to
control the gauge degrees of freedom of the canonical states.
Equations \eqref{Cb-TDHFB-1}, \eqref{Cb-TDHFB-2}, and \eqref{Cb-TDHFB-3}
are invariant with respect to the gauge transformation with
arbitrary real functions, $\theta_i(t)$ and $\theta_{\bar i}(t)$.
\begin{eqnarray}
\label{gauge_transf_1}
\ket{\phi_i}\rightarrow e^{i\theta_i(t)}\ket{\phi_i}
\quad &\mbox{and}& \quad
\ket{\phi_{\bar i}}\rightarrow e^{i\theta_{\bar i}(t)}\ket{\phi_{\bar i}}
\\
\label{gauge_transf_2}
\kappa_k\rightarrow e^{-i(\theta_i(t)+\theta_{\bar i}(t))}\kappa_i
\quad &\mbox{and}& \quad
\Delta_k\rightarrow e^{-i(\theta_i(t)+\theta_{\bar i}(t))}\Delta_i
\end{eqnarray}
simultaneously with
\begin{equation}
\eta_i(t)\rightarrow \eta_i(t)+\frac{d\theta_i}{dt}
\quad \mbox{and} \quad
\eta_{\bar i}(t)\rightarrow \eta_{\bar i}(t)+\frac{d\theta_{\bar i}}{dt} .
\end{equation}
Thus, $\eta_i(t)$ and $\eta_{\bar i}(t)$,
control time evolution of the phases of
$\ket{\phi_i(t)}$, $\ket{\phi_{\bar i}(t)}$, $\kappa_i(t)$, and
$\Delta_i(t)$.

The canonical-basis TDHFB equations \eqref{Cb-TDHFB-1},
\eqref{Cb-TDHFB-2}, and \eqref{Cb-TDHFB-3},
with a proper gauge choice
 guarantee the following properties \cite{Eba10}:
\begin{enumerate}
\item Conservation law
 \begin{enumerate}
 \item Conservation of the orthonormal property of the canonical states
 \item Conservation of the average particle number
 \item Conservation of the average total energy
 \end{enumerate}
\item The stationary solution corresponds to the HF+BCS state.
\item In the small-amplitude limit,
 the Nambu-Goldstone modes correspond to zero-energy normal-mode solutions.
\end{enumerate}

\begin{wrapfigure}{r}{0.41\textwidth}
\vspace{-15pt}
\begin{flushright}
\includegraphics[height=0.40\textwidth,angle=-90]{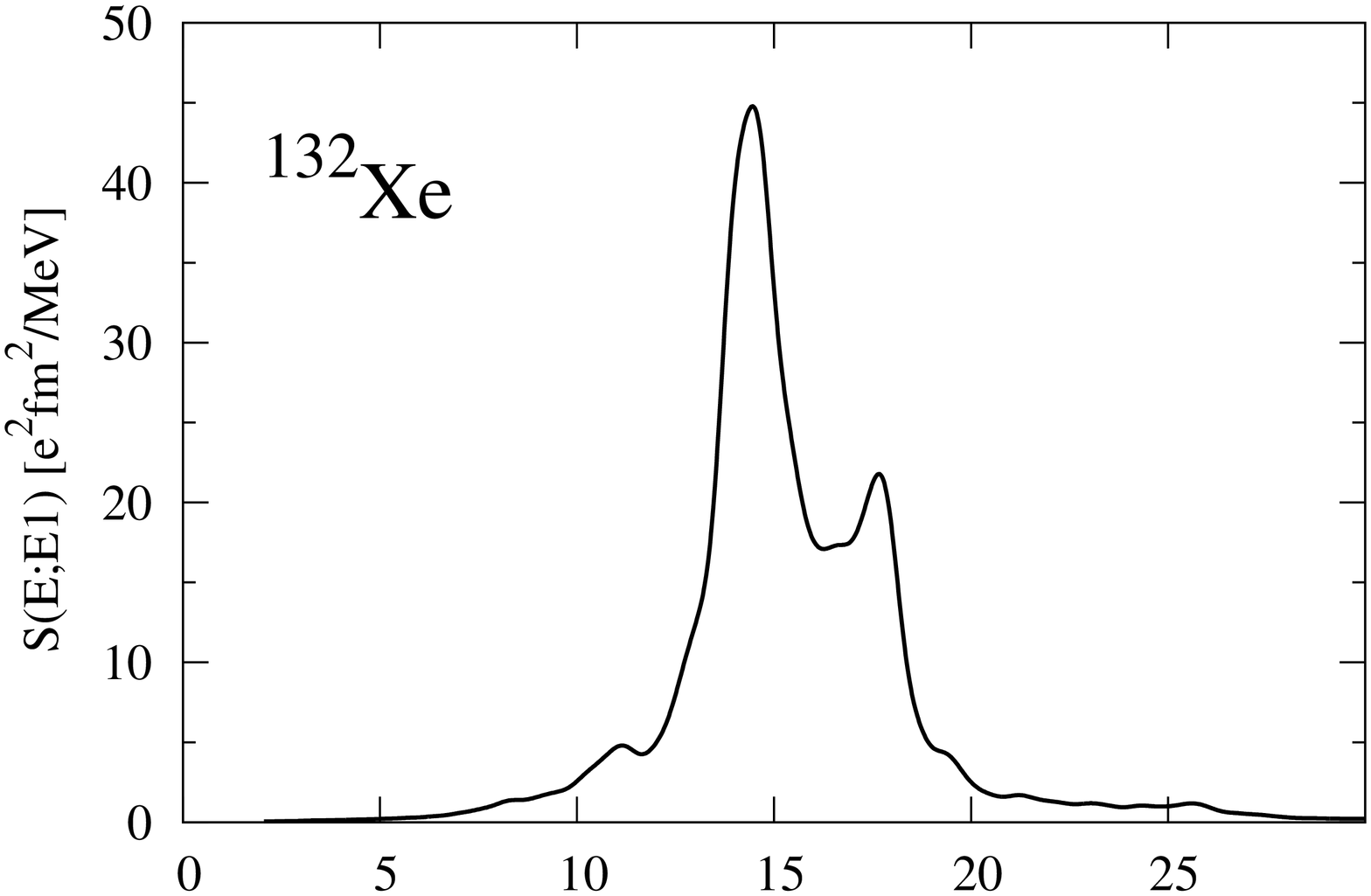}\newline
\includegraphics[height=0.40\textwidth,angle=-90]{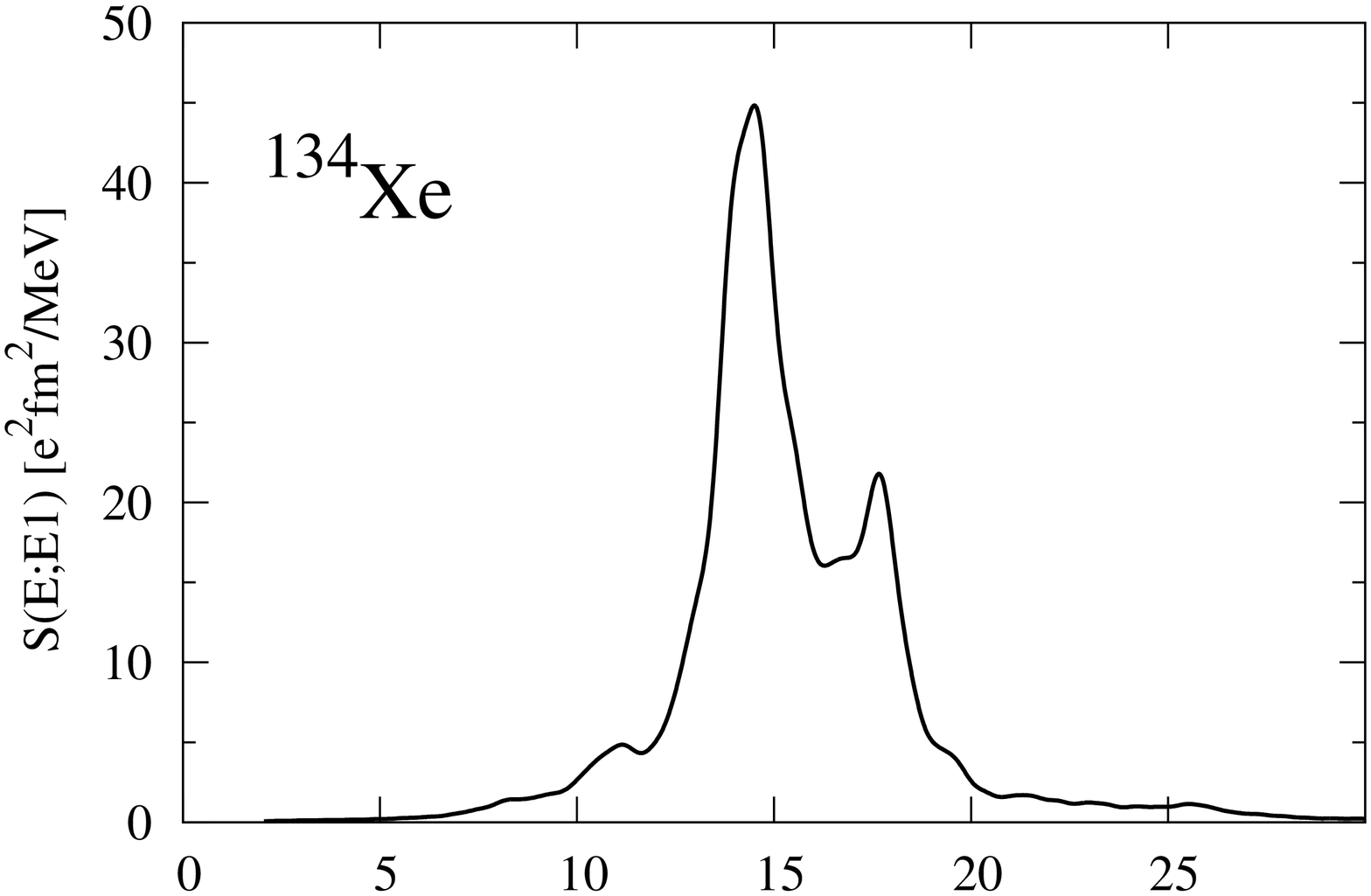}\newline
\includegraphics[height=0.40\textwidth,angle=-90]{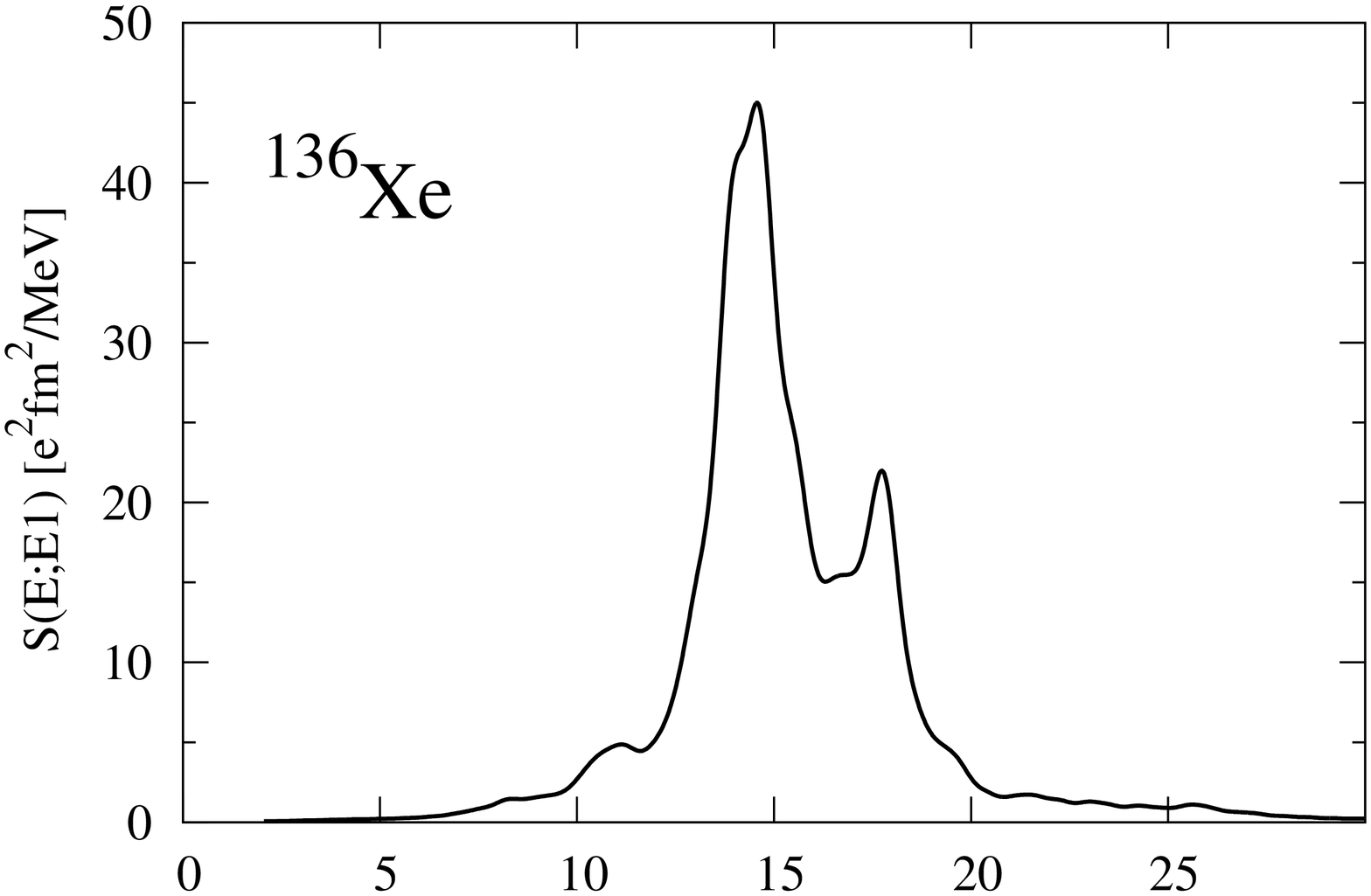}\newline
\includegraphics[height=0.40\textwidth,angle=-90]{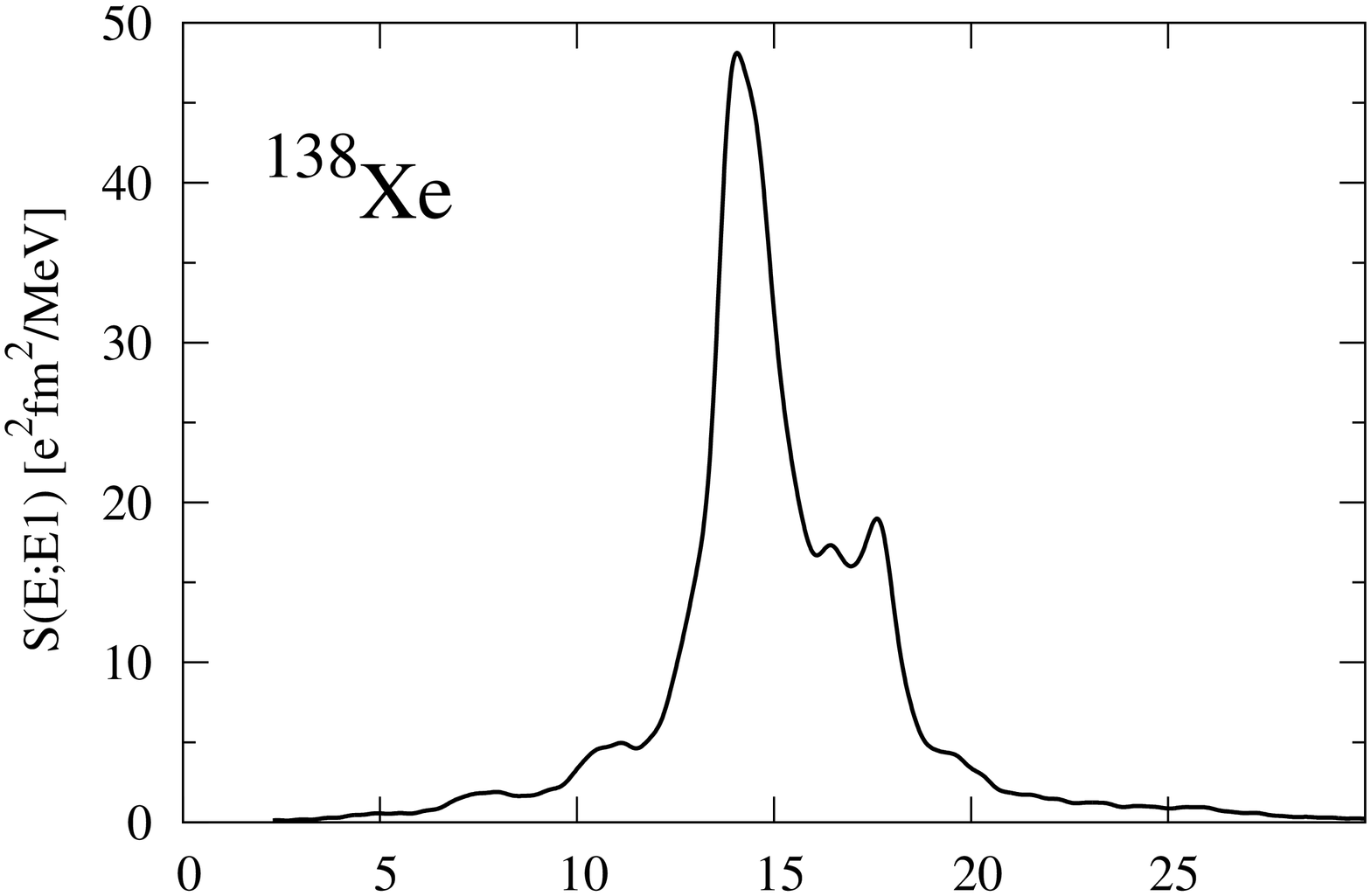}\newline
\includegraphics[height=0.40\textwidth,angle=-90]{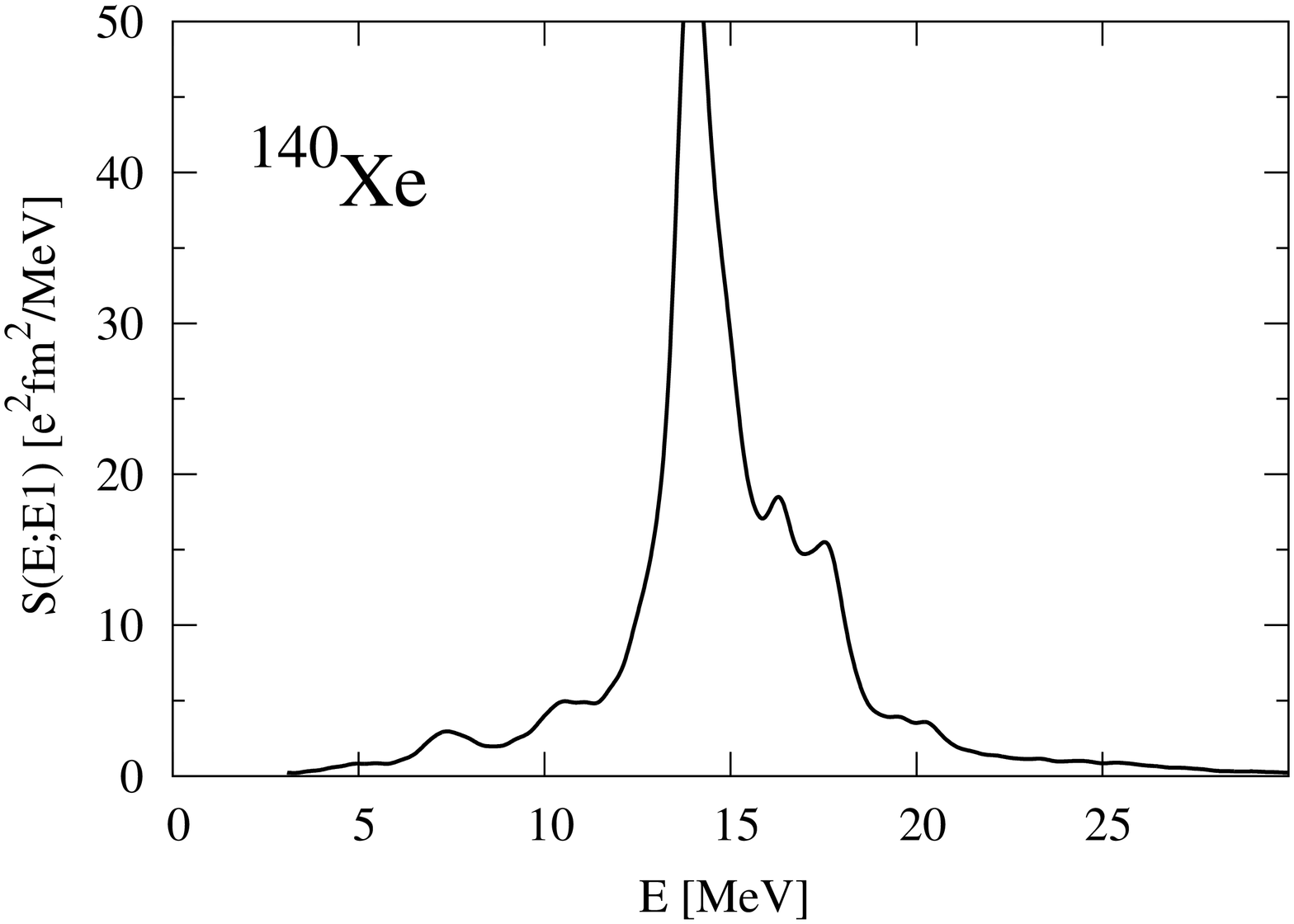}
\end{flushright}
\caption{$E1$ strength distributions in Xe isotopes
calculated with the SkM* functional.
}
\label{fig: Xe}
\end{wrapfigure}

In this study, although the method is applicable to the
large amplitude dynamics, it is utilized to study the photoreaction
of Xe isotopes in the linear regime.
Using the instantaneous perturbative external $E1$ field,
we calculate the time evolution in the 3D grid-space representation.
Then, the Fourier transform of the time-dependent $E1$ moment leads
to the $E1$ response function.
The smoothing parameter of $\Gamma=1$ MeV is used.
See Ref.~\cite{Eba10} for numerical details.
The calculated $E1$ strength distributions in $^{132-140}$Xe are shown
in Fig.~\ref{fig: Xe}.

The ground states of these nuclei are calculated to be spherical,
except for $^{140}$Xe which has a very small deformation of $\beta=0.02$.
The peak energy of the giant dipole resonance (GDR) is roughly constant
and $E_{\rm GDR}\approx 15$ MeV for these isotopes.
$^{136}$Xe corresponds to the neutron magic number $N=82$.
The property of the GDR does not significantly depend on the neutron magicity.
In the low-energy region below $E=10$ MeV, we may notice an onset of
a small dipole peak beyond $N=82$, which appear in $^{138}$Xe and
increases in $^{140}$Xe.
This seems to be mainly due to a drastic decrease in the neutron 
separation energy beyond $N=82$.
A systematic study on the low-energy $E1$ strength in this mass region
is currently under progress.

It should be emphasized that the computational cost of the present
real-time approach is significantly smaller than the normal QRPA
calculation based on the diagonalization method.
For instance, the computational cost of the QRPA calculation
in Ref. \cite{TE10} is larger than the present one by several orders
of magnitude, even though the axial symmetry restriction was utilized.
This is because the canonical-basis method utilize the selected
canonical states whose number is much smaller than the number of
qp states.
In addition, the real-time method is very efficient for the present
purpose, because the single time evolution produces the nuclear response
for the entire energy region.

\section{Summary}

We presented basic concepts of the density functional theory in
nuclear physics, based on the historical developments in nuclear
many-body theories.
The energy density functional is established as the
density-dependent effective interactions,
which gives a consistent mean-field-type description of the nuclear
saturation property.
The time-dependent density functional approach is a powerful
method to study dynamics of the quantum many-body systems.
In description of heavy open-shell nuclei,
the Kohn-Sham orbitals should be extended to
the Bogoliubov-type quasiparticle orbitals,
to include the pair correlations.
This is essentially identical to the TDHFB theory.
Although the full treatment of the TDHFB is still a
challenging task, we presented approximated treatments;
the finite amplitude method and the canonical-basis TDHFB method.
Both methods serve as an efficient computational approach to
dynamical properties of heavy superfluid nuclei.

\section*{Acknowledgements}
This work was supported by Grant-in-Aid for Scientific
Research(B) No. 21340073 and Innovative Areas No. 20105003.
The numerical calculations were performed in part on the
RIKEN Integrated Cluster of Clusters (RICC), 
on PACS-CS and T2K in Center for Computational Sciences,
University of Tsukuba, and
on HITACHI SR16000 in Yukawa Institute, Kyoto University.

\end{document}